\documentclass[12pt]{iopart}

\begin{document}

\title[Analogue gravity accelerated observers in a BEC]{Accelerated observers emerging from a Bose--Einstein condensate through analogue gravity}

\author{B Gonz\'alez-Fern\'andez$^1$ and A Camacho$^2$}

\address{$^1$ Departamento de Ciencias e Ingenier\'ias, Universidad Iberoamericana Puebla\\ Blvrd. del Ni\~no Poblano 2901, Reserva Territorial Atlixc\'ayotl, 72820 San Andr\'es Cholula, Pue., M\'exico}

\address{$^2$ Departamento de F\'{\i}sica,
 Universidad Aut\'onoma Metropolitana--Iztapalapa\\
 Apartado Postal 55--534, C.P. 09340, M\'exico, D.F., M\'exico}

\eads{\mailto{belinka.gonzalez.fernandez@iberopuebla.mx}, \mailto{acq@xanum.uam.mx}}

\date{\today}

\begin{abstract}
The current article explores the problem of a Bose--Einstein condensate (BEC) trapped in an anisotropic three-dimensional harmonic-oscillator potential, with positive effective interatomic interactions, and the analogue gravity acoustic metric emerging from it. We find that the latter gives rise to a family of conformally equivalent metrics and the space--time interval corresponds to that of an accelerated observer subject to a position--dependent acceleration. Furthermore, such acceleration can be deduced differentiating the potential of the trap, which leads us to infer that, for a Bose--Einstein condensate in an arbitrary trap, the space--time interval will correspond to the conformal equivalent of that of an accelerated observer, being the acceleration in any direction proportional to the partial derivative of the trap potential of the BEC with respect to that coordinate.  Finally, we compute the space--time interval, the Einstein tensor and the geodesic equations, identifying some of the Killing vectors, for three cases: the spherically symmetric, the axially symmetric and the asymmetric traps.
\end{abstract}


\submitto{\CQG}

\maketitle


\section{Introduction}

Analogies play a fundamental role in physics and
mathematics, as they define a bridge between two different
branches of science. One of the current trends in this direction is
the so--called Analogue Gravity, a topic that considers the
similarities between physics in a curved space-time and, usually,
hydrodynamic systems \cite{Barc1}. 

Even when the existence of an analogy among these models does not imply a
complete equivalence (there is still no analogue system that reproduces Einstein's equations \cite{Volo1}), it can capture a sufficient number of relevant common features so that it is possible to use the known science in one of them to study phenomena of the other.

One of the simplest cases of such approach involves a moving fluid and an analogue
space-time provided by the acoustics of the fluid \cite{Visser1,Misner1}.
The analysis of the speed of sound in a fluid and the possibility of
shaping it in the form of the  equation of motion for a
non--massive field immersed in a curved space-time is already an old
result \cite{Barc1}, which leads to an equation of motion for the
corresponding velocity potential that can be expressed as an equation of motion for a minimally--coupled scalar field, without mass, which 
propagates in a Lorentzian geometry. Some 
generalizations of this result, such as the possibility of having a
rotational fluid, \cite{Unruh1,Garcia1,LiberatiV16,LiberatiV18}, or the case of an irrotational viscous fluid \cite{BelinkaCamacho1}, have also been carried out.

In particular, we are interested in the search for analogies between condensed matter and gravitational cases, for they could shed light upon some major problems in gravity. 
Bose--Einstein condensates (BEC) stand out for being conceptually clear and well--understood experimentally--controlled systems, which appear to be great candidates to simulate several phenomena predicted in High Energy Physics \cite{Barcelo2003,Carusotto2008,Penrose,Garay}.  For instance, since condensed matter physics is described resorting to a quantum
field theory which can be tested experimentally, the analysis of
the analogies between particle physics, particularly the quantum
vacuum, and condensed matter could clarify the physics for
trans--Planckian situations \cite{Volo1}.

Thus, interesting deviations from the hydrodynamic or relativistic regimes in BEC could show the way to the kinematic corrections that a quantum gravity theory might impose on general relativity, allowing us to study quantum gravity phenomenology by means of the analogy.

In the present work, we study the analogue gravity acoustic metric emerging from a Bose--Einstein condensate, with interatomic interactions, trapped in an anisotropic three-dimensional harmonic-oscillator potential.  In section \ref{section2} we revisit the formulation for an emerging analogue acoustic metric for a BEC and, then, focus on that trapped in a harmonic-oscillator potential to, eventually, discuss the physical interpretation of the space--time element.  Section \ref{section3} is dedicated to the study of the gravitational quantities and dynamics of the system in three cases: the spherically--symmetric, the axially--symmetric and the asymmetric traps.  Finally, in section \ref{section4} we discuss our results.

\section{Bose--Einstein Condensates and Their Analogy to Gravity}\label{section2}

\subsection{Emerging acoustic metric for a BEC}\label{subsection2.1}

The derivation of the acoustic metric for a BEC system has been developed before \cite{Barc1}, showing that the equations of the phonons of the condensate mimic the dynamics of a scalar field in a curved space-time.

It is useful to remember that in the Thomas--Fermi approximation \cite{Pethick1}, any gas under an external potential $ V_{ext} $ can be described in terms of a quantum field $ \Psi $, satisfying the equation 

\begin{equation}
i \hbar\frac{\partial}{\partial t}\hat{\Psi} =  \left( -\frac{\hbar^{2}}{2m}\nabla^{2}+V_{ext}+\kappa(a)\hat{\Psi}^{\dag} \hat{\Psi} \right) \hat{\Psi}  ,
\label{blv226}
\end{equation}

where we assume that the effective interaction between the particles is parametrized by $ \kappa(a) $.  In our case, in which the energies are very low and the wavelength excitations are very long, we can express it in terms of the scattering length $ a $ as

\begin{equation}
\kappa(a) = \frac{4 \pi a \hbar^{2}}{m} ,
\label{blv227}
\end{equation}

being $ m $ the mass of one of the particles that constitute the BEC.

It has already been established, \cite{Barc1}, that separating the quantum field in its classical part (macroscopic condensate) and the fluctuations, for the case in which the back-reaction can be neglected for an irrotational velocity field, the equations of motion of the system can be identified with those of an inviscid irrotational fluid.  Thus, in the acoustic representation, the equations of motion for the field describing the quantum perturbations of the system, $\hat{\theta}_{1}$, can be cast such as that of a massless minimally coupled quantum scalar field over a curved background, so that
so that the equation of motion for it can be written as

\begin{equation}
\Delta \hat{\theta}_{1} \equiv \frac{1}{\sqrt{-g}} \partial_{\mu}(\sqrt{-g}g^{\mu\nu}\partial_{\nu}\hat{\theta}_{1}) = 0.
\label{blv254}
\end{equation}

In this case, the effective metric is

\begin{equation}
g_{\mu\nu}(t, \bi{r}) \equiv \frac{n_{c}}{m c_{s}}    
\left( \begin{array}{ccc}
-\left[c_{s}^{2} - v^{2} \right]  & \vdots & -v_{j}\\
\cdots &  & \cdots \\
 -v_{i} & \vdots & \delta_{ij}
\end{array} \right)   ,
 \label{blv246}
\end{equation}

where the speed of the phonons in the medium is represented by 

\begin{equation}
c_{s}\equiv c_{s}(a, n_{c})^{2} = \kappa(a) n_{c}/m
\label{blv256}
\end{equation}

and 
\begin{equation}
n_{c} \equiv \vert \psi(t, \textbf{r}) \vert^{2}.
\label{blv231}
\end{equation}

\subsection{Acoustic metric for a BEC in a harmonic-oscillator trap}\label{subsection2.2}

The system we are to analyse is a dilute Bose--Einstein condensate confined in an anisotropic three-dimensional harmonic-oscillator trap, under repulsive interatomic interactions (which produce no stability problems). For such a system, subject to the three-dimensional harmonic-oscillator, time-independent, potential

\begin{equation}
V(x, y, z) = \frac{m}{2} \left( \omega_{x}^{2} x^{2} + \omega_{y}^{2} y^{2} + \omega_{z}^{2} z^{2} \right)  ,
\label{p6.16}
\end{equation}

the wave function of the condensate is

\begin{equation}
\psi (\textbf{r}) = \left( \frac{N}{\pi^{3/2}bcd} \right)^{1/2} \
e^{-\frac{1}{2} \left( \frac{x^{2}}{b^{2}} + \frac{y^{2}}{c^{2}}+\frac{z^{2}}{d^{2}}\right)} , 
\label{p6.18}
\end{equation}

where $b, c$ and $d$ are the variational parameters in each direction, \cite{Pethick1}.

Substituting (\ref{p6.18}) in (\ref{blv231}) we find

\begin{equation}
n_{c}(\textbf{r}) = \frac{N}{\pi^{3/4}bcd}e^{- \left( \frac{x^{2}}{b^{2}} + \frac{y^{2}}{c^{2}}+\frac{z^{2}}{d^{2}}\right)},
\label{i1}
\end{equation}

which is time-independent and using the values given by equations (\ref{blv227}) and (\ref{i1}) in (\ref{blv256}),

\begin{equation}
c_{s}^{2} = K e^{- \left( \frac{x^{2}}{b^{2}} + \frac{y^{2}}{c^{2}}+\frac{z^{2}}{d^{2}}\right)},
\label{vsc}
\end{equation}

where, for the sake of simplicity, we have defined

\begin{equation}
K \equiv \frac{4 N \hbar^{2} a}{\pi^{1/2} m^{2} bcd }.
\label{K}
\end{equation}

In turn, the coefficient multiplying of the matrix in (\ref{blv254}) is

\begin{equation}
\frac{n_{c}}{m c_{s}} = \alpha e^{-\frac{1}{2} \left( \frac{x^{2}}{b^{2}} + \frac{y^{2}}{c^{2}}+\frac{z^{2}}{d^{2}}\right)} ,
\label{ai38}
\end{equation}

being

\begin{equation}
\alpha \equiv \left[\frac{N}{4 \pi^{5/2} \hbar^{2}  bcd a }\right]^{1/2}.
\label{A}
\end{equation}

Substituting (\ref{vsc}) and (\ref{ai38}) in (\ref{blv246}), we find the effective acoustic metric for this system to be

\begin{eqnarray}
 g_{\mu\nu} = \alpha e^{-\frac{1}{2} \left( \frac{x^{2}}{b^{2}} + \frac{y^{2}}{c^{2}}+\frac{z^{2}}{d^{2}}\right)} 
 \left( \begin{array}{ccc}
 -\left[ K e^{- \left( \frac{x^{2}}{b^{2}} + \frac{y^{2}}{c^{2}}+\frac{z^{2}}{d^{2}}\right)} - v^{2} \right]  & \vdots & -v_{j}\\
   \cdots &  & \cdots \\
    -v_{i} & \vdots & \delta_{ij}
  \end{array} \right)   .
  \label{r1}
\end{eqnarray}

According to Bose--Einstein condensation physics \cite{Ueda}, for this system $\textbf{v}=0$, so that (\ref{r1}) is reduced to

\begin{eqnarray}
 g_{\mu\nu} = \alpha e^{-\frac{1}{2} \left( \frac{x^{2}}{b^{2}} + \frac{y^{2}}{c^{2}}+\frac{z^{2}}{d^{2}}\right)} 
 \left( \begin{array}{ccc}
 -K e^{- \left( \frac{x^{2}}{b^{2}} + \frac{y^{2}}{c^{2}}+\frac{z^{2}}{d^{2}}\right)} & \vdots & 0\\
   \cdots &  & \cdots \\
   0 & \vdots & \delta_{ij}
  \end{array} \right)   .
  \label{gtg}
\end{eqnarray}

We can identify the global factor of the matrix in (\ref{gtg}) as a conformal one, so that this expression describes a family of conformally equivalent metrics.  As it has been discussed in literature, \cite{Plebanksi}, the fundamental geometric properties of this kind of system can be deduced setting the conformal factor equal to unity, so that (\ref{gtg}) can be expressed as

\begin{eqnarray}
 g_{\mu\nu} = 
 \left( \begin{array}{ccc}
 -K e^{- \left( \frac{x^{2}}{b^{2}} + \frac{y^{2}}{c^{2}}+\frac{z^{2}}{d^{2}}\right)} & \vdots & 0\\
   \cdots &  & \cdots \\
   0 & \vdots & \delta_{ij}
  \end{array} \right)   ,
  \label{gtgs}
\end{eqnarray}

and the space-time interval will be

\begin{equation}
ds^{2}=
 \left[ -K {\rm e}^{-\left( {\frac {x^{2}}{{b}^{2}}} + \frac {y^{2}}{c^{2}} + \frac {z^{2}}{d^{2}}\right)} dt^{2}+ dx^{2}+ dy^2 + d z^2 \right].
\label{elsfc}
\end{equation}

\subsection{Physical interpretation of the space--time element}\label{subsection2.3}

In the context of general relativity \cite{Misner1}, it has been established that the line interval for an observer being accelerated in the $ \xi^{1}$ direction on a Minkowski space--time is

\begin{equation}
ds^2 \equiv -(1+g \xi^1)^2(d \xi^0)^2+(d \xi^1)^2+(d \xi^2)^2+(d \xi^3)^2, 
\label{mtw6.18}
\end{equation}

where the $  \xi^{\mu} $ are the coordinates of the system relative to the observer and $g$ is its acceleration.  Such coordinate system approximates a Lorentz one in its immediate neighbourhood, \textit{i.e.} only in the region in which the condition

\begin{equation}
g \xi^{1} \ll 1
\label{cl}
\end{equation}
 
is satisfied, so that we can neglect the quadratic term in the binomial factor in (\ref{mtw6.18}) and write

\begin{equation}
ds^2 \simeq -(1+2g \xi^1)(d \xi^0)^2+(d \xi^1)^2+(d \xi^2)^2+(d \xi^3)^2. 
\label{mtw6.18a}
\end{equation}

Generalizing this result for a situation in which the observer moves with an arbitrary 3--acceleration, we find

\begin{equation}
ds^2 \simeq -(1+2g_{i} \xi^i)(d \xi^0)^2+(d \xi^1)^2+(d \xi^2)^2+(d \xi^3)^2, 
\label{mtw6.18g}
\end{equation}

where we have defined the spatial components of the acceleration and the coordinates of the local system as $g^{i}$ and $\xi^{i}$, respectively ($i = 1, 2, 3$).  

To contrast the last equation with the line interval for the BEC in the trap, (\ref{elsfc}), 
we first rescale the time-coordinate so that its differential will read $d t' \equiv \sqrt{K} dt$, expand the exponential and keep only the quadratic terms in the spatial coordinates so that we obtain

\begin{equation}
ds^{2}\simeq -
 \left[ 1-\left( {\frac {x^{2}}{{b}^{2}}} + \frac {y^{2}}{c^{2}} + \frac {z^{2}}{d^{2}}\right)\right] dt'^{2}+ dx^{2}+ dy^2 + d z^2 .
\label{ela}
\end{equation}

To verify this approximation is compatible with (\ref{cl}), it suffices to notice that the points inside the trap, which define our analogue-manifold always satisfy
\begin{eqnarray}
\frac{x_{i}}{b_{i}} < 1 & \Rightarrow & \frac{x_{i}^{2}}{b_{i}^{2}} \ll 1 ,
\label{clbec}
\end{eqnarray}
\\
so that the condition of locality is fulfilled by the analogue system (being $x_{1}\equiv x$, $x_{2}\equiv y$ and $x_{3}\equiv z$, while $b_{1}\equiv b$, $b_{2}\equiv c$ and $b_{3}\equiv d$).

When we re-write (\ref{ela}) as

\begin{equation}
ds^{2}=-
 \left[ 1-\left({\frac {x}{{b}^{2}}}x  + \frac {y}{c^{2}}y  +\frac {z}{d^{2}}z\right)\right] dt'^{2}+ dx^{2}+ dy^2 + d z^2 ,
\label{elac}
\end{equation}
and compare (\ref{mtw6.18g}) and (\ref{elac}), defining the correspondence of the local coordinates $(\xi^{0}, \xi^{1}, \xi^{2}, \xi^{3})$ with $(t', x, y, z)$, we conclude that the analogue--gravitational system to that of a BEC in a harmonic--oscillator trap can be regarded as an accelerated observer, whose acceleration components are

\begin{equation}
g_{i} \equiv -\frac{x_{i}}{2 b_{i}^{2}}.
\label{aobec}
\end{equation}

It is worth noting that the acceleration is position--dependent and negative, so that the analogue observer would be accelerated towards the origin of the coordinate system, that is center of the trap, and the closer to the boundary, the larger the induced force would be.

Recalling expression (\ref{p6.16}), it is possible to define a potential per mass unit 

\begin{equation}
V' \equiv \frac{V}{m}= \frac{1}{2} \left( \omega_{x}^{2} x^{2} + \omega_{y}^{2} y^{2} + \omega_{z}^{2} z^{2} \right)  ,
\label{p6.16d}
\end{equation}

so that the acceleration of a particle inside a harmonic--oscillator trap, along the direction $x_{i}$, can be found by differentiating (\ref{p6.16d}) with respect to that coordinate

\begin{equation}
a_{i}= - \frac{\partial V'}{\partial x_{i}} = - \omega_{i}^{2} x\textit{i} .
\label{pa}
\end{equation} 

Contrasting (\ref{pa}) and (\ref{aobec}), it follows that the acceleration of the analogue observer will be the partial derivative of the trap potential of the BEC with respect to that coordinate, with $\omega_{i}^{2}=(2b_{i}^{2})^{-1}$, all these expressions are written in geometrized units.

Thus, we conjecture that in the most general case, that is, the gravitational--analogue of a BEC trapped by an arbitrary potential $V(x, y, z)$, the space--time interval will correspond to the conformal equivalent of that of an accelerated observer and the acceleration experienced in any direction will be proportional to the partial derivative of the trap potential of the BEC with respect to that coordinate. In particular, this would imply that for a BEC in a box type-like potential there would be no acceleration of the analogue--observer, as $V=0$ inside the condensate.

\section{Gravitational quantities and dynamics of the system}\label{section3}

For completeness, in this section we analyse the three possible cases of symmetry in this system, computing the space--time interval, the Einstein tensor and the geodesic equations for each case, along with some Killing vectors.

\subsection{Spherically symmetric trap}\label{subsection3.1}

In the case of maximal symmetry, all the semi-axes of the ellipse have the same length, thus $b=c=d$.  Under these circumstances, constants (\ref{K}) and (\ref{A}) reduce to 

\begin{equation}
K_{s} = \frac{4 N \hbar^{2} a}{\pi^{1/2} m^{2} b^{3} }
\label{K1}
\end{equation}

and

\begin{equation}
\alpha_{s} = \left[\frac{N}{4 \pi^{5/2} \hbar^{2} b^{3} a }\right]^{1/2}.
\label{A1}
\end{equation}

Due to the symmetry of the system, it is most convenient to use spherical coordinates to study this case, so that the metric can be written as

\begin{eqnarray}
 g_{\mu\nu}^s = \alpha_{s} {\rm e}^{-\frac {{r}^{2}}{2{b}^{2}}}
 \left( \begin{array}{ccc}
 -K_{s} {\rm e}^{-{\frac {{r}^{2}}{{b}^{2}}}}& \vdots & 0\\
   \cdots &  & \cdots \\
   0 & \vdots & \delta_{ij}
  \end{array} \right)   
  \label{gts}
\end{eqnarray}

and the space--time interval is

\begin{equation}
ds_{s}^{2}= \alpha_{s} {\rm e}^{-\frac {{r}^{2}}{2{b}^{2}}}
 \left( -K_{s} {\rm e}^{-{\frac {{r}^{2}}{{b}^{2}}}} dt^{2}+ dr^{2}+ r^2 d\theta^2 + r^2 sin^2\theta d\phi^2 \right).
\label{elts}
\end{equation}

With this information, we computed the Einstein tensor

\begin{eqnarray}
\fl
G_{\mu\nu}^s = 
\left( \begin{array}{cccc}
\frac {K_{s} \left( 12{b}^{2}-{r}^{2} \right)}{4{b}^{4}}{
{\rm e}^{-{\frac {{r}^{2}}{{b}^{2}}}}}
 & 0 & 0 & 0\\
 0 & -\frac {16{b}^{2}-7{r}^{2}}{4{b}^{4}} & 0 & 0 \\
   0 & 0 & -\frac {{r}^{2} \left( 16{b}^{2}-9{r}^{2}\right)}{4{b}^{4}}& 0 \\
0 & 0 & 0 & -\frac {{r}^{2} \sin^{2}\theta  \left( 16{b}^{2}-9{r}^{2} \right) }{4{b}^{4}}
  \end{array} \right)   
  \label{ets}
\end{eqnarray}

while the geodesics turned out to be

\numparts

\begin{eqnarray}
\fl
\frac{d^2t}{d\lambda ^2} - \frac{3r}{b^2} \left( \frac{dt}{d\lambda}\right) \left(\frac{dr}{d\lambda} \right)  =0 ,
\label{gets}
\\ \nonumber
\\ \nonumber
\\  \nonumber
\fl
\frac{d^2r}{d\lambda ^2}-\frac {3K_{s}r}{2{b}^{2}}{{\rm e}^{-{\frac {{r}^{2}}{{b}^{2}}}}}\left( \frac{dt}{d\lambda}\right)^2 -\frac {r}{2{b}^{2}}\left( \frac{dr}{d\lambda}\right)^2 
-\frac {r \left( 2{b}^{2}-{r}^{2} \right) }{2{b}^{2}}\left( \frac{d\theta}{d\lambda}\right)^2 
\\ 
-\frac {r \sin ^2 \theta \left( 2{b}^{2}-{r}^{2} \right) }{2{b}^{2}}\left( \frac{d\phi}{d\lambda}\right)^2=0 ,
\label{gers}
\\ \nonumber
\\ \nonumber
\\ 
\fl
\frac{d^2\theta}{d\lambda ^2}+ \frac {2{b}^{2}-{r}^{2}}{r{b}^{2}} \left(  \frac{dr}{d\lambda} \right) \left(  \frac{d\theta}{d\lambda}\right)  - \sin\theta \cos\theta\left( \frac{d\phi}{d\lambda}\right)^2 = 0 ,
\label{gethetas} 
\\ \nonumber
\\ \nonumber
\\ 
\fl
\frac{d^2\phi}{d\lambda ^2}+ \frac {2{b}^{2}-{r}^{2}}{r{b}^{2}} \left( \frac{dr}{d\lambda} \right) \left(  \frac{d\phi}{d\lambda}\right) + 2 \cot \theta \left(  \frac{d\theta}{d\lambda} \right) \left(  \frac{d\phi}{d\lambda}\right)=0.
\label{gephis}
\end{eqnarray}
\endnumparts
\\
Finally, as the metric in this case is independent of the time and the $\phi$ angle, there are, at least, two conserved quantities: energy and axial angular momentum.  This, in turn, implies that there are two Killing vectors associated with these quantities \cite{Misner1} which we proved to be $\zeta_{1}^{s}= (1,0,0,0)$ and $\zeta_{2}^{s}= (0,0,0,1)$ in this coordinate system. 

\subsection{Axially symmetric trap}\label{subsection3.2}

When two of the semi-axes are equal, we face the case with axial symmetry, in which $b=c$, but $d$ is different.  Thus, in this case the constants (\ref{K}) and (\ref{A}) are

\begin{equation}
K_{c} = \frac{4 N \hbar^{2} a}{\pi^{1/2} m^{2} b^{2} d }
\label{K2}
\end{equation}

and

\begin{equation}
\alpha_{c} = \left[\frac{N}{4 \pi^{5/2} \hbar^{2} b^{2} d a }\right]^{1/2}.
\label{A2}
\end{equation}

For this case, we chose to use cylindrical coordinates, so that the metric reads

\begin{eqnarray}
 g_{\mu\nu}^c = \alpha_{c} {\rm e}^{-\frac{1}{2} \left( \frac {r^{2}}{b^{2}} + \frac {z^{2}}{d^{2}} \right) }
 \left( \begin{array}{cccc}
-K_{c} {\rm e}^{-\left( {\frac {{r}^{2}}{{b}^{2}}} + \frac {z^{2}}{d^{2}}\right)}& 0 & 0 & 0\\
  0 & 1 & 0 & 0\\
   0 & 0 & r^{2} & 0 \\
   0 & 0 & 0 & 1
  \end{array} \right)   
  \label{gsaz}
\end{eqnarray}

and the space--time interval,

\begin{equation}
ds_{c}^{2}= \alpha_{c} {\rm e}^{-\frac{1}{2} \left( \frac {r^{2}}{b^{2}} + \frac {z^{2}}{d^{2}} \right) }
 \left[ -K_{s} {\rm e}^{-\left( {\frac {{r}^{2}}{{b}^{2}}} + \frac {z^{2}}{d^{2}}\right)} dt^{2}+ dr^{2}+ r^2 d\theta^2 + d z^2 \right].
\label{elsaz}
\end{equation}

In turn, the Einstein tensor for the system is

\begin{eqnarray}
\fl 
G_{\mu\nu}^c = 
  \nonumber\\
\fl
\left( \begin{array}{cccc}
K_{c} \left( -{\frac {{r}^{2}}{{b}^{4}}}-{\frac {{z}^{2}}{{d}^{4}}}+\frac{8}{b^{2}}+\frac{4}{{d}^{2}} \right) 
{\rm e}^{-\left( {\frac {{r}^{2}}{{b}^{2}}} + \frac {z^{2}}{d^{2}}\right)}
 & 0 & 0 & 0\\
 0 & {\frac {7{r}^{2}}{4{b}^{4}}}+{\frac {9{z}^{2}}{4{d}^{4}}}-\frac{2}{b^{2}}-\frac{2}{d^{2}}
 & 0 & -{\frac {rz}{2{b}^{2}{d}^{2}}}  \\
 0 & 0 &  r^2 \left( {\frac {{r}^{2}}{{b}^{4}}}+{\frac {{z}^{2}}{{d}^{4}}}-\frac{3}{b^2}-\frac{1}{2d^{2}} \right)  & 0 \\
0 & -{\frac {rz}{2{b}^{2}{d}^{2}}} & 0 & \frac {9{r}^{2}}{4{b}^{4}}+\frac {7{z}^{2}}{4{d}^{4}}-\frac{4}{b^2}
\end{array} \right)   ,
 \nonumber\\
\label{etsaz}
\end{eqnarray}

which reduces to the case in \ref{subsection3.1} when we set $b \equiv d$.
\\

The geodesics for this symmetry case are

\numparts

\begin{eqnarray}
\fl
\frac{d^2t}{d\lambda ^2} - \frac{3r}{b^2} \left( \frac{dt}{d\lambda}\right) \left(\frac{dr}{d\lambda} \right) - \frac{3z}{d^2} \left( \frac{dt}{d\lambda}\right) \left(\frac{dz}{d\lambda} \right) =0 ,
\label{gtsaz}\\ \nonumber
\\ \nonumber
\\ \nonumber
\fl
\frac{d^2r}{d\lambda ^2}   
-\frac {3K_{c}r}{2{b}^{2}}{ {\rm e}^{ -\left( {\frac {{r}^{2}}{{b}^{2}}}+{\frac {{z}^{2}}{{d}^{2}}}\right)  }}\left( \frac{dt}{d\lambda}\right)^2 
-\frac {r}{2{b}^{2}}\left[ \left( \frac{dr}{d\lambda}\right)^2 + \left(\frac{dz}{d\lambda} \right)^2  \right]
\\
-\frac {r \left( 2{b}^{2}-{r}^{2} \right) }{2{b}^{2}}\left( \frac{d\theta}{d\lambda}\right)^2 
-\frac{z}{d^2}\left(\frac{dr}{d\lambda} \right)\left(\frac{dz}{d\lambda} \right)=0 ,
\label{grsaz}
\\ \nonumber
\\ \nonumber
\\ 
\fl
\frac{d^2\theta}{d\lambda ^2}+ 
\frac {2{b}^{2}-{r}^{2}}{r{b}^{2}} \left(\frac{dr}{d\lambda} \right) \left(\frac{d\theta}{d\lambda}\right)  
- \frac{z}{d^2} \left( \frac{dr}{d\lambda} \right) \left( \frac{d\theta}{dz}\right) = 0 ,
\label{gthetasaz} 
\\ \nonumber
\\ \nonumber
\\
\fl
\frac{d^2z}{d\lambda ^2}   
-\frac {3K_{c}z}{2{d}^{2}}{ {\rm e}^{ -\left( {\frac {{r}^{2}}{{b}^{2}}}+{\frac {{z}^{2}}{{d}^{2}}}\right)  }}\left( \frac{dt}{d\lambda}\right)^2 
+\frac {z}{2{d}^{2}}\left[ \left( \frac{dr}{d\lambda}\right)^2 - \left(\frac{dz}{d\lambda} \right)^2  \right]
\\ \nonumber
-\frac {z {r}^{2}}{2{d}^{2}}\left( \frac{d\theta}{d\lambda}\right)^2 
-\frac{r}{b^2}\left(\frac{dr}{d\lambda} \right)\left(\frac{dz}{d\lambda} \right)=0 .
\label{gzsaz}
\end{eqnarray}
\endnumparts
\\

In this case, the metric in this case is also independent of the time and the $\phi$ angle, so that energy and axial angular momentum are conserved quantities in this system too.  The two Killing vectors associated with these quantities, in cylindrical coordinates, are $\zeta_{1}^{c}= (1,0,0,0)$ and $\zeta_{2}^{c}= (0,0,1,0)$.

\subsection{Asymmetric trap}\label{subsection3.3}

At last, we analyse the case in which none of the semi-axes are equal. The constants (\ref{K}) and (\ref{A}) for the asymmetric case are

\begin{equation}
K_{a} = \frac{4 N \hbar^{2} a}{\pi^{1/2} m^{2} (b c d) }
\label{K3}
\end{equation}

and

\begin{equation}
\alpha_{a} = \left[\frac{N}{4 \pi^{5/2} \hbar^{2} a(b c d)  }\right]^{1/2}.
\label{A3}
\end{equation}

The most convenient choice of coordinates for the asymmetry is Cartesian, in which the metric is

\begin{eqnarray}
 g_{\mu\nu}^a = \alpha_{a} {\rm e}^{-\frac{1}{2} \left( \frac {x^{2}}{b^{2}} + \frac {y^{2}}{c^{2}} + \frac {z^{2}}{d^{2}} \right) }
\left( \begin{array}{ccc}
 -K_{a} {\rm e}^{- \left( \frac{x^{2}}{b^{2}} + \frac{y^{2}}{c^{2}}+\frac{z^{2}}{d^{2}}\right)} & \vdots & 0\\
   \cdots &  & \cdots \\
   0 & \vdots & \delta_{ij}
  \end{array} \right) 
  \label{gtasc}
\end{eqnarray}

so, the space--time interval is

\begin{equation}
ds_{a}^{2}= \alpha_{c} {\rm e}^{-\frac{1}{2} \left( \frac {x^{2}}{b^{2}} + \frac {y^{2}}{c^{2}} + \frac {z^{2}}{d^{2}} \right) }
 \left[ -K_{a} {\rm e}^{-\left( {\frac {x^{2}}{{b}^{2}}} + \frac {y^{2}}{c^{2}} + \frac {z^{2}}{d^{2}}\right)} dt^{2}+ dx^{2}+ dy^2 + d z^2 \right].
\label{elasc}
\end{equation}

Here, the the Einstein tensor is

\begin{eqnarray}
\fl 
G_{\mu\nu}^a = 
  \nonumber\\
\fl
\left( \begin{array}{cccc}

G_{00}^a
 & 0 & 0 & 0\\
 0 & \frac {7x^{2}}{4b^{4}} + \frac {9y^{2}}{4c^{4}} + \frac {9z^{2}}{4d^{4}}-\frac{2}{c^2} -\frac{2}{d^2}
 & -{\frac {xy}{2{b}^{2}{c}^{2}}} & -{\frac {xz}{2{b}^{2}{d}^{2}}}  \\
0 & -{\frac {xy}{2{b}^{2}{c}^{2}}} & \frac {9x^{2}}{4b^{4}} + \frac {7y^{2}}{4c^{4}} + \frac {9z^{2}}{4d^{4}}-\frac{2}{b^2} -\frac{2}{d^2} & -{\frac {yz}{2{c}^{2}{b}^{2}}}\\
0 & -{\frac {xz}{2{b}^{2}{d}^{2}}} & -{\frac {yz}{2{c}^{2}{b}^{2}}} & \frac {9x^{2}}{4b^{4}} + \frac {9y^{2}}{4c^{4}} + \frac {7z^{2}}{4d^{4}}-\frac{2}{b^2} -\frac{2}{c^2}
  \end{array} \right)   ,
  \nonumber\\
  \label{etasc}
\end{eqnarray}

where we have defined

\begin{equation}
\fl
G_{00}^a = K_{a}\left[ -\frac{1}{4} \left( \frac {x^{2}}{b^{4}} + \frac {y^{2}}{c^{4}} + \frac {z^{2}}{d^{4}} \right)
+ \frac{1}{b^2}+ \frac{1}{c^2}+ \frac{1}{d^2} \right] 
{\rm e}^{-\left( {\frac {{x}^{2}}{{b}^{2}}} + \frac {y^{2}}{c^{2}} + \frac {z^{2}}{d^{2}}\right)} .
\end{equation}

As expected, the latter reduces to the cases in \ref{subsection3.2} and \ref{subsection3.1} when the appropriate coordinate transformations are made and constants are set as they should. 
\\

The geodesics for this symmetry case are

\numparts

\begin{eqnarray}
\fl
\frac{d^2t}{d\lambda ^2} - 3 \left( \frac{dt}{d\lambda}\right)\left[  \frac{x}{b^2} \left(\frac{dx}{d\lambda} \right) +\frac{y}{c^2} \left(\frac{dy}{d\lambda} \right)-\frac{z}{d^2} \left(\frac{dz}{d\lambda} \right) \right] = 0 , 
\label{gat}\\ \nonumber
\\ \nonumber
\\ \nonumber
\fl
\frac{d^2 x}{d\lambda ^2}   
-\frac {3K_{a}x}{2{b}^{2}}{ {\rm e}^{ -\left( {\frac {{x}^{2}}{{b}^{2}}} + \frac {y^{2}}{c^{2}} +{\frac {{z}^{2}}{{d}^{2}}}\right)  }}\left( \frac{dt}{d\lambda}\right)^2 
+ \frac{x}{2b^{2}}\left[-\left(\frac{dx}{d\lambda}\right)^2 + \left( \frac{dy}{d\lambda}\right)^2 + \left(\frac{dz}{d\lambda} \right)^2 \right]
\\ 
-\left( \frac{dx}{d\lambda}\right) \left[ \frac{y}{c^2}\left(\frac{dy}{d\lambda}\right)  + \frac{z}{d^2}\left(\frac{dz}{d\lambda}\right) \right] = 0 ,
\label{gax}
\\ \nonumber
\\ \nonumber
\\ \nonumber
\fl
\frac{d^2 y}{d\lambda ^2}   
-\frac {3K_{a}y}{2{c}^{2}}{ {\rm e}^{ -\left( {\frac {{x}^{2}}{{b}^{2}}} + \frac {y^{2}}{c^{2}} +{\frac {{z}^{2}}{{d}^{2}}}\right)  }}\left( \frac{dt}{d\lambda}\right)^2 
+ \frac{y}{2c^{2}}\left[\left(\frac{dx}{d\lambda}\right)^2 - \left( \frac{dy}{d\lambda}\right)^2 + \left(\frac{dz}{d\lambda} \right)^2 \right]
\\ 
-\left( \frac{dy}{d\lambda}\right) \left[ \frac{x}{b^2}\left(\frac{dx}{d\lambda}\right)  + \frac{z}{d^2}\left(\frac{dz}{d\lambda}\right) \right] = 0 ,
\label{gay} 
\\ \nonumber
\\ \nonumber
\\ \nonumber
\fl
\frac{d^2 z}{d\lambda ^2}   
-\frac {3K_{a}z}{2{d}^{2}}{ {\rm e}^{ -\left( {\frac {{x}^{2}}{{b}^{2}}} + \frac {y^{2}}{c^{2}} +{\frac {{z}^{2}}{{d}^{2}}}\right)  }}\left( \frac{dt}{d\lambda}\right)^2 
+ \frac{z}{2d^{2}}\left[\left(\frac{dx}{d\lambda}\right)^2 + \left( \frac{dy}{d\lambda}\right)^2 - \left(\frac{dz}{d\lambda} \right)^2 \right]
\\ 
-\left( \frac{dz}{d\lambda}\right) \left[ \frac{x}{b^2}\left(\frac{dx}{d\lambda}\right)  + \frac{y}{c^2}\left(\frac{dy}{d\lambda}\right) \right] = 0.
\label{gaz}
\end{eqnarray}
\endnumparts
\\

In the asymmetric case, the only conserved quantity is energy, and its associated killing vector in Cartesian coordinates is $\zeta^{a}= (1,0,0,0)$.


\section{Conclusions}\label{section4}

In this work, we have studied the gravitational analogue emerging from a Bose--Einstein condensate trapped in an anisotropic three-dimensional harmonic-oscillator potential, with positive effective interatomic interactions, finding that the analogue gravity acoustic metric of the system gives rise to a family of conformally equivalent metrics.

By comparing the condensate system with its general--relativity counterpart, we realised that
the analogue space--time interval corresponds to that of an accelerated observer subject to a position--dependent acceleration, from which we conclude that the gravitational analogue to a BEC in a trap is consistent with that of a differential manifold, in which accelerated observers emerge. 

We also found that such acceleration can be obtained differentiating the potential of the trap with respect to the spatial coordinates.  From this, we come to the conclusion that, for a Bose--Einstein condensate in an arbitrary trap, the space--time interval will correspond to the conformal equivalent of that for an accelerated observer, in which the acceleration along the $x_{i}$ direction will be proportional to $-\partial V / \partial x_{i}$.

Finally, for completeness, we computed the space--time interval, the Einstein tensor and the geodesic equations for the three possible kinds of traps: the spherically symmetric, the axially symmetric and the asymmetric ones.  We proved that each case reduces to the others under the corresponding symmetry assumptions.  We also found that the only conserved quantity in every case is energy, whereas the axial angular momentum is conserved only in the spherically and axially symmetric traps, and found the Killing vectors corresponding to such quantities in the three cases.

\ack

B. Gonz\'alez--Fern\'andez wants to acknowledge Universidad Iberoamericana Puebla for the support given to this project.


\section*{References}

\begin{thebibliography}{21}

\bibitem{Barc1}
Barcel\'o C, Liberati S and Visser M 2011 Analogue Gravity \textit{Living Reviews in Relativity} \textbf{14}

\bibitem{Volo1}
Volovik  G E 2002 \textit{The Universe in a Helium Droplet} (Oxford: Oxford Science Publications)


\bibitem{Visser1}
Visser M 1998, \textit{Clas. Quant. Grav.} \textbf{15}, 1767--1791

\bibitem{Misner1}
Misner C, Thorne  K and Wheeler J A 1973 \textit{Gravitation} (San Francisco: W. H. Freeman)

\bibitem{Unruh1}
Unruh W G 1981 Experimental Black-Hole Evaporation? \textit{Phys. Rev. Lett.}, \textbf{46(21)}, 1351–1353
https://doi.org/10.1103/PhysRevLett.46.1351

\bibitem{Garcia1}
Garcia de Andrade L C  2004 \textit{Phys. Rev.} \textbf{D70}, 064004; 2005
\textit{Phys. Lett.} \textbf{A339}, 188--193 ; 2005 \textit{Phys. Lett.}
\textbf{A346}, 327--329

\bibitem{LiberatiV16}
Cropp B, Liberati S and Turcati R 2016 Vorticity in analogue gravity \textit{Classical and Quantum Gravity} \textbf{33(12)} 125009
https://doi.org/10.1088/0264-9381/33/12/125009

\bibitem{LiberatiV18}
Liberati S, Schuster S, Tricella G and Visser M 2018. \textit{Vorticity in analogue space-times}. Retrieved from http://arxiv.org/abs/1802.04785

\bibitem{BelinkaCamacho1}
Gonz\'alez--Fern\'andez B and Camacho A 2012, Fluid--Gravity Correspondence Under the Presence of Viscosity \textit{Modern Physics Letters A} \textbf{27(32),} 1250185 https://doi.org/10.1142/S0217732312501854

\bibitem{Barcelo2003}
Barcel\'o C, Liberati S, and Visser M 2003 Towards the Observation of Hawking Radiation in Bose–Einstein Condensates \textit{International Journal of Modern Physics A} \textbf{18(21)} 3735–3745 https://doi.org/10.1142/S0217751X0301615X

\bibitem{Carusotto2008}
Carusotto I, Fagnocchi S, Recati A, Balbinot R and Fabbri A 2008 Numerical observation of Hawking radiation from acoustic black holes in atomic Bose–Einstein condensates \textit{New Journal of Physics} \textbf{10(10)}
https://doi.org/10.1088/1367-2630/10/10/103001

\bibitem{Penrose}
Howl R, Penrose R and Fuentes I 2018 \textit{Exploring the unification of quantum theory and general relativity with a Bose-Einstein condensate} Retrieved from http://arxiv.org/abs/1812.04630

\bibitem{Garay} 
Garay L, Anglin J, Cirac J and Zoller P 2000 Sonic Analog of Gravitational Black Holes in Bose-Einstein Condensates\textit{ Physical Review Letters} \textbf{85(22)} 4643–4647 https://doi.org/10.1103/PhysRevLett.85.4643


\bibitem{Pethick1}
 Pethick C J and Smith H 2004 {\it Bose--Einstein Condensation in Dilute Gases} (Cambridge: Cambridge University Press)

\bibitem{Ueda}
Ueda M 2010 \textit{Fundamentals and New Frontiers of Bose-Einstein Condensation} World Scientific
https://doi.org/10.1142/7216

\bibitem{Chorin}
Chorin A L and Marsden J E 2000 \textit{A Mathematical Introduction to
Fluid Mechanics} (New York, Springer--Verlag)

\bibitem{Holmes}
Holmes M H 1998 \textit{Introduction to Perturbation Methods} (New York, Springer--Verlag)

\bibitem{Pitaevski1}
Pitaevski L P and Stringari S 2003 \textit{Bose--Einstein Condensation} (Oxford, Oxford University Press)

\bibitem{Patterson} 
Patterson J and Bailey B 2010 \textit{Solid-State Physics: Introduction to the Theory} (Springer)
http://doi.org/10.1007/978-3-642-02589-1

\bibitem{Pathria1}
 Pathria R K 1996 {\it Statistical Mechanics} (Oxford: Butterworth--Heinemann)

\bibitem{Plebanksi}
Pleba\'nski J and Krasi\'nski A 2006 \textit{An introduction to general relativity and cosmology}. (Cambridge University Press)


\end{thebibliography}
\end{document}